\documentclass[journal]{IEEEtran}
\ifCLASSINFOpdf
\else
\fi
\usepackage{cite}
\usepackage{graphicx,epstopdf}
\usepackage{textcomp}
\usepackage{mathtools}
\usepackage{multirow}
\usepackage{adjustbox}
\usepackage{tabularx}
\usepackage[table,xcdraw]{xcolor}
\usepackage{xcolor}
\usepackage{amsmath}
\usepackage{soul}

\begin{document}
%
\title{\textit{mmPose-NLP}: A Natural Language Processing Approach to Precise Skeletal Pose Estimation using mmWave Radars}
%
%
%

\author{\IEEEauthorblockN{\large{Arindam Sengupta and Siyang Cao}}\\
\IEEEauthorblockA{{Department of Electrical and Computer Engineering, University of Arizona, Tucson, AZ USA} \\
Email: \{sengupta, caos\}@email.arizona.edu}
}

\maketitle

\begin{abstract}
\boldmath
In this paper we presented \textit{mmPose-NLP}, a novel Natural Language Processing (NLP) inspired Sequence-to-Sequence (Seq2Seq) skeletal key-point estimator using millimeter-wave (mmWave) radar data. To the best of the author's knowledge, this is the first method to precisely estimate upto 25 skeletal key-points using mmWave radar data alone. Skeletal pose estimation is critical in several applications ranging from autonomous vehicles, traffic monitoring, patient monitoring, gait analysis, to defense security forensics, and aid both preventative and actionable decision making. The use of mmWave radars for this task, over traditionally employed optical sensors, provide several advantages, primarily its operational robustness to scene lighting and adverse weather conditions, where optical sensor performance degrade significantly. The mmWave radar point-cloud (PCL) data is first voxelized (analogous to tokenization in NLP) and $N$ frames of the voxelized radar data (analogous to a text paragraph in NLP) is subjected to the proposed \textit{mmPose-NLP} architecture, where the voxel indices of the 25 skeletal key-points (analogous to keyword extraction in NLP) are predicted. The voxel indices are converted back to real world 3-D coordinates using the voxel dictionary used during the tokenization process. Mean Absolute Error (MAE) metrics were used to measure the accuracy of the proposed system against the ground truth, with the proposed \textit{mmPose-NLP} offering $<$3~cm localization errors in the depth, horizontal and vertical axes. The effect of the number of input frames vs performance/accuracy was also studied for $N = \{1,2,\dots,10\}$. A comprehensive methodology, results, discussions and limitations are presented in this paper. All the source codes and results are made available on GitHub for furthering research and development in this critical yet emerging domain of skeletal key-point estimation using mmWave radars.  
\end{abstract}


%
\IEEEpeerreviewmaketitle

%
%
%
%

\section{Introduction}
\IEEEPARstart{A}{rtifical} Intelligence aided sensing systems use a combination of Machine Learning (ML) and Computer Vision (CV) approaches to perceive the environment. Besides well established schemes to identify objects in the scene, one of the emerging areas of CV based applications is that of skeletal key-point or pose estimation, i.e identifying specific joints of the human body from still images or video data. This can be used to provide granular information of the detected human target in terms of the posture/pose, which can be extremely critical for several applications. Skeletal pose could provide traffic monitoring systems and autonomous vehicles additional information about the current state of a pedestrian, and predict their upcoming motion (walk, run, jump etc.), which in turn aids in preventative decision making. Pose estimation can also find applications in automated remote patient monitoring, given the current dearth in healthcare monitoring staff \cite{oulton2006global}. Furthermore, the skeletal key-point kinematics can find applications in Gait analysis, both from a Bio-mechanics clinical setting, and also in forensic analysis in a defense security and surveillance setting \cite{gait1,gait2,gait3,gait4}.

\par Current skeletal pose estimation is predominantly carried out with the aid of optical sensors. However, optical sensors suffer operationally in poor lighting, occlusion, and adverse weather conditions such as rain, fog or snow. This heavy reliance on optical sensors for perception has led to several autonomous car crashes, on account of the sensor failing due to inadequate scene lighting or over-exposure, even resulting in a pedestrian's loss of life \cite{Uber, Tesla}. Furthermore, one of the biggest challenges in using optical sensors in patient monitoring systems is the ever increasing concerns for user privacy. On the other hand, current gait analysis, and several patient monitoring systems make use of biometric information using wearable sensors, in some cases for prolonged amounts of time, which is uncomfortable and require constant maintenance, and does not translate to a long-term commitment according to a recent public survey \cite{pwc}.

\par Millimeter-Wave (mmWave) radars on the other hand addresses most of these critical challenges. First and foremost, mmWave radars use their own RF signals to illuminate the target, thereby making them operationally robust to external scene lighting and weather conditions. The radar signals captures the target's silhouette via a sparse 3-D point cloud (PCL) representation without capturing facial features of the users, thereby complying with their privacy. Finally, they are low-cost, low-power, and compact to serve as a non-invasive, non-wearable monitoring solution, making them practical to deploy. Infact, autonomous vehicles do carry mmWave radars among its myriad of sensors, but are not used to its full potential and are limited to the role of a secondary sensor, mainly due to its inferior resolution than the optical counterparts. Furthermore, limited availability of annotated radar data-sets make research and development all the more challenging.  

\par In this paper, we present \textit{mmPose-NLP}, a skeletal key-point estimation approach using mmWave radar signals, inspired from an abstractive text summarization application in Natural Language Processing (NLP). To the best of our knowledge, this is the first approach to estimate the 3-D positions of 25 skeletal keypoints using mmWave radar data alone. Continuing our pursuit to develop accurate mmWave radar based skeletal key-point estimation techniques that started with \textit{mm-Pose}\cite{mmPose}, this work is a logical extension of our previous simulation based study \cite{mmPosenlp}. The rest of the paper is organized as follows. An overview of existing skeletal pose estimation techniques is reviewed in Section.~\ref{literature}. Relevant background theory for radar and the building blocks of \textit{mmPose-NLP} is presented in Section.~\ref{BT}. The proposed approach is presented in Section.~\ref{proposed1}, followed by results, discussion and limitations of the proposed approach in Section.~\ref{results}. Finally, the paper is summarized and concluded in Section.~\ref{conclusion}.

\section{Related Work}\label{literature}
Skeletal pose estimation has been predominantly explored in the realm of optical sensors dominated CV community, using a combination of image processing and machine learning. One of the early works in 2005 was \textit{Strike a Pose}, that used rectangular templates to localize and identify 10 distict body section from still images and videos \cite{ramanan2005strike}. A key-point based multi-person detection scheme was proposed in 2016 using $k$-poselet based agglomerative clustering \cite{gkioxari2014using}. A ResNet-RCNN based scheme was proposed in 2017 to estimate $N$ skeletal key-points by learning $N$ mask templates \cite{he2017mask}. In 2016, \textit{DeepCut} and a faster-and-accurate \textit{DeeperCut} were proposed, that used a bottom-up approach to multi-person pose estimation using ResNets  \cite{insafutdinov2016deepercut, pishchulin2016deepcut}. Alternately, a top-down approach was explored by Google by first using an R-CNN to localize regions of interest, and then a ResNet architecture to estimate the skeletal key-points, achieving the then-highest precision metric on the the {Common Objects in Context (COCO)} test-dev set \cite{papandreou2017towards}. One of the most popular real-time skeletal pose estimation algorithms to this day is Carnegie Mellon University's \textit{OpenPose}, proposed in 2017 \cite{cao2017realtime}. \textit{OpenPose} used a bottom-up approach by using Part Affinity Fields (PAFs), essentially a non-parametric representation of different body parts, and won the COCO key-points challenge in 2016. The open-source data-sets and compatibility across various platforms make \textit{OpenPose} one of the most popular benchmarks, are are also used to generate ground-truth skeletal key-point data to train skeletal pose estimators using a non-optical sensing modality. 

\par The aforementioned approaches primarily use data from monocular still or video cameras and therefore lack the depth information, which is typically non-trivial and challenging to extract. The skeletal key-points, therefore are 2-D in nature, represented by the pixel co-ordinates. In order to extend skeletal key-point estimation to real-word 3-D coordinate system, University of Toronto created the \textit{HumanEva} dataset, by using a circular array of 7 video cameras capturing the scene \cite{sigal2010humaneva}. The ground truth was obtained using a commercial ViconPeak motion capture system that used reflective markers attached to specific joint locations of the human subjects. Besides using multiple video cameras to estimate 3-D key-points, Microsoft Kinect uses an infra-red (IR) sensor, that estimates depth in tandem with a per-pixel classification aided RGB camera to estimate skeletal key-point using a probabilistic approach \cite{zhang2012microsoft}. Despite significant developments in pose estimation techniques using optical sensors (camera/IR), their inability to operate in poor lighting, occlusion and the increasing concerns for privacy significantly limits their use in several critical applications. Therefore, alternate sensing modalities are are also being explored for the skeletal pose estimation task, including RF sensors such as WiFi and radars.

\begin{figure*}[th!]
\centering
\includegraphics[scale=1]{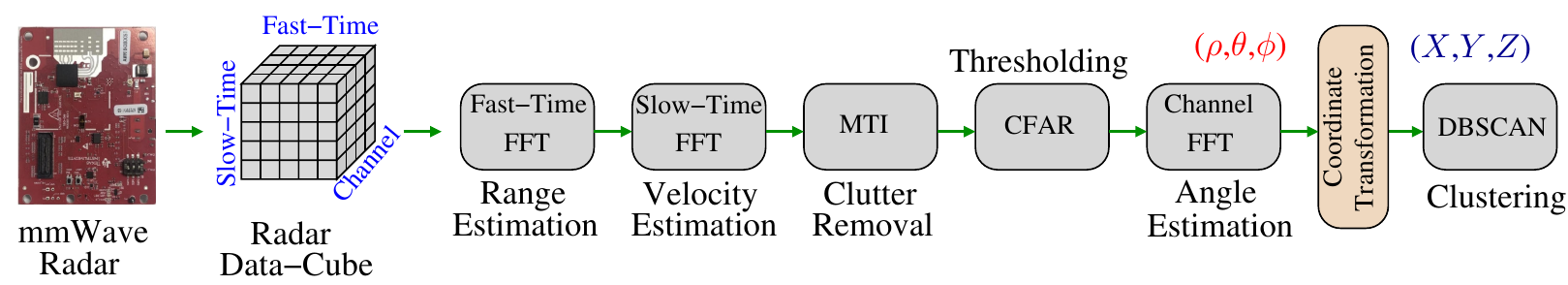}
\caption{The Radar Signal Processing Chain.}
\vspace{-0.5cm}
\label{RSPC}
\end{figure*}

\par Although several behavior prediction approaches using radar signals have been proposed previously, primarily using the motion parameters of the reflection points, the skeletal key-points in 3-D space were not explicitly estimated \cite{ref_patientbehaviour,leobehavior,mmfall}. Skeleton key-point estimation using RF signals is a relatively new and emerging area of research limited by a lower resolution and unavailability of public datasets. Pioneered by MIT's \textit{CSAIL Lab}, the first set of radar based pose estimation techniques came into existence, the first being \textit{RF-Capture}, in 2015, that used Frequency-Modulated Continuous Wave (FMCW) signals and an antenna array to independently identify several parts of the human body and combine them in post-processing to reconstruct a human figure \cite{Adib:2015:CHF:2816795.2818072}. This was followed by \textit{RF-Pose} in 2018, that used RF heat-maps obtained using a vertical and a horizontal antenna arrays, and an encoder-decoder architecture was used to estimate the skeletal key-points\cite{Zhao_2018_CVPR}. Another approach by the same group, \textit{RF-Based 3D Skeletons}, used a 1.8 GHz wide FMCW signalsto estimate the 3-D positions of key-points, using a ResNet architecture, followed by triangulation to estimate a 3-D model of a human skeleton with 8 skeletal key-points. A circular array of vision sensors was used to capture the scene and Open-Pose skeletal data served as the ground truth for supervised training \cite{RFpose}. In 2020, we proposed mm-Pose, a realtime skeletal pose estimation approach using 77 GHz mmWave radar, that used a novel low-complexity representation of the radar point-cloud and reflected power levels as input to a forked CNN architecture to estimate 17 skeletal keypoints \cite{mmPose}. A through-wall pose imaging using a {3.3 - 10} GHz FMCW antenna array was proposed to estimate 15 skeletal key-points with a combination of convolutional neural network (CNN), region proposal network and recurrent neural networks (RNNs) \cite{8999267}. A heat-map based pose estimation using two 77 GHz mmWave radars was proposed by \textit{Li et. al.}, in a similar setup to our previously proposed \textit{mm-Pose}, however, a discussion around the achieved skeletal key-point localization errors was not covered \cite{newkeypoint}. Finally, a proof-of-concept NLP inspired Seq2Seq skeletal keypoint estimation using simulated radar point-cloud data from Kinect acquired 3-D skeletal key-point data was presented \cite{mmPosenlp}. However, the study did not include validation using actual radar PCL data. 
\par In this paper, we present \textit{mmPose-NLP}, inspired from an abstractive text summarization application in NLP, that uses $N$ consecutive frames of radar reflection data, and summarizes it to the desired 25 skeletal key-points. This work is an extension of \cite{mmPosenlp}, however with a change to the input data structure. As opposed to \cite{mmPosenlp}, where we concatenated two frames of radar data into a single tensor, the proposed approach sends $N$ radar frames sequentially. Furthermore, the proposed approach also studies the localization accuracy trends for various values of $N$. In summary, our major contributions include: 1) the first method to estimate 25 skeletal key-points using mmWave data; 2) lowest localization errors among mmWave radar based skeletal key-point estimation techniques, and 3) the study of localization accuracy with respect to the number of input frames $N$. 

\section{Background Theory}\label{BT}
\subsection{\color{black}{Radar Signal Processing Chain}}\label{RSPC-sec}
Radars are time-of-flight sensors that use their own RF signals to illuminate the target, and use the reflected signals to estimate the target's range and velocity. The radar essentially sends out signal with a bandwidth $BW$, in pulses separated by a pulse-repetition-interval ($PRI$). The range resolution ($\Delta R$) is a function of $BW$ and is given as $\Delta R = \frac{c}{2BW}$, while the maximum unambiguous range $R_{ua}$ can be expressed as a function of $PRI$ as $R_{ua} = \frac{cPRI}{2}$, where $c$ is the speed of light. Rectangular pulse's width, or in other words power, is inversely proportional to the BW, which implies that the power carried by a pulse to achieve a fine range resolution would be very low, which in turn would result in the received reflected signal to be very faint to be detected from noise. 
\par To overcome the power-bandwidth tradeoff, Linear Frequency Modulated (LFM) signals use pulse compression to offer a high-bandwidth, while retaining enough width to maintain the necessary energy budget. A frame of radar data constitues radar returns from $N$ ``chirps'' over a coherent-processing-interval ($CPI = N\times PRI$), which determines the velocity or doppler resolution. During acquisition, every chirp return is subjected to an analog-to-digital converter that converts it to $M$ discretized samples, referred to as the fast-time axis. In a frame or CPI, this process is repeated for $N$ consecutive chirps, resulting in an $N\times M$ matrix, where the other axis is referred to as the slow-time axis. If the reciever has $L$ antenna channels, each sensing element would generate this detection matrix, resulting in s 3-D $N \times M \times L$ radar data-cube with fast-time, slow-time and channel axes.

\par The Radar Signal Processing Chain (RSPC) \cite{rsp} starts with the raw 3-D radar data-cube. First, Fast Fourier Transform (FFT) along the fast-time axis is performed, where the beat-freaquency is proportional to the time-delay between the transmitted and recieved signal, which in turn helps to determine the range ($\rho$) of the target(s). An FFT along the slow-time axis aids in determining the doppler frequencies, which is used to compute the radial velocity ($v$) of the target(s). Moving Target Indicator (MTI) is used to isolate the desired targets from static clutter by virtue of delay lines \cite{MTI}. For a more reliable detection, Constant False Alarm Rate (CFAR) is used to estimate the noise floor around the target, and avoid false detections by setting a detection SNR threshold \cite{CFAR}. Finally, an FFT along the channel dimension helps determine the azimuth ($\theta$) and elevation ($\phi$) angles of the targets. The resulting output from this RSPC is $(\rho,\theta, \phi, v)$ of all the detected targets. The output in spherical co-ordinate system can be converted to a 3-D ($X,Y,Z$) cartesian coordinate system using:
\begin{equation}
\begin{split}
X &= \rho\sin\phi\cos\theta\\
Y &= \rho\sin\phi\sin\theta\\
Z &= \rho\cos\phi\\
\end{split}
\label{RSPC-eqn}
\end{equation}

\par mmWave radars operate at 76-81~GHz, with the transmit chirps upto 4~GHz in bandwidth. This results in range resolutions of the order of 3-5~cm. Therefore a single target may have multiple points of reflection. In order to cluster reflection points from the same object a Density Based Spatial Clustering of Applications with Noise (DBSCAN) is used \cite{DBSCAN}. The overall RSPC is depicted in Fig.~\ref{RSPC}. Furthermore typical mmWave radar transceivers for automotive applications carry multiple transmit $(Tx_1, Tx_2 \dots ,Tx_{n_t})$ and receive $(Rx_1, Rx_2 \dots ,Rx_{n_r})$ channels. A Time-Division-Multiplexing (TDM) scheme is employed to successively send out a chirp, with the carrier wavelength $\lambda$, from a transmit channel $Tx_i$, $\forall i \in [1,\dots,n_t]$, and the reflected signals are sensed by all the $n_r$ array of receiving channels. Based on the received signal's direction of arrival $\theta$, and inter-element spacing $d$, the receiver channels receive the echo with a progressive phase term \(\Psi (n)\), corresponding to the path/time delay in receiving the signal across multiple channels. For the $n^{th}$ reciever element $Rx_n$, $n \in [1,\dots,n_r]$, we have:
\begin{equation}
\Psi(n) = \frac{2\pi*n*d*\sin\theta}{\lambda}.
\label{equ_AOA}
\end{equation}
\par The angular resolution is governed by the antenna size, in this case, $(n_r-1)\times d$. However, drive towards compact sensing solutions results in limited real-estate that can facilitate only a few antenna elements. Using multiple-input multiple-output (MIMO) principles, it is possible to increase the antenna size/channels ``virtually". One way this can be achieved is with the inter-element spacing on the transmitting channels to be set at $n_r \times d$. With this setting, we get a virtual receiver antenna size of $(n_r\times n_t - 1)*d$ as opposed to $(n_r-1)\times d$, thereby improving the angular resolution without having to fabricate additional antenna channels. A similar TDM-MIMO approach can also be used to estimate and resolve for the direction of arrival in the elevation direction ($\phi$), and a similar detailed discussion has been omitted for redundancy and brevity.             
\subsection{\color{black}{Sequence-to-Sequence Models}}
With the increase in computation resources and power, Neural Networks (NN) have emerged as one of the most popular machine learning (ML) techniques to cater to a wide range of classification and regression applications. Loosely derived from biological neurons, NNs are an inter-connection of neurons or nodes, that are essentially computational units whose basic function is to take in a weighted input, and apply a non-linear transformation to it. These nodes are densely interconnected and are organized in several layers or sub-stages. In a fully connected NN, every node $i$ in $j^{th}$ layer is connected to all the $k=1,2, \dots , N$ nodes in $(j-1)^{th}$ layer, with each connection scaling the outputs $O_k$ from the previous nodes by a weight $W_{i,k}$ (the bias terms are absorbed in the weight representation, here). The cumulative sum of these inputs at node $i$ is then subjected to a non-linear activation function $\Omega(.)$. The intermediate layers in an ANN projects the input data to a higher-dimensional feature space where they are separable, while the activation function aids with determining the non-linear boundaries of separation. 
\par At the final layer $f$, the outputs $O_{f,i} = \Omega(\sum\limits_{k=1}^{N} W_{i,k} \times O_k)$ are then used to compare against the desired output, and a suitable loss function is used to determine the performance of the NN. The weights across the entire ANN are readjusted to reduce the loss, using back-propagation aided gradient descent or its variants. Therefore, with an extensive set of labeled input-output pairs, sufficient depth (layers) and nodes in the networks, and an appropriate choice of activation function and loss functions, NNs ``learn" the optimum values of weights required to achieve our desired applications. With these fundamental principles intact, several application specific variants such as Convolutional Neural Networks (CNNs) and Recurrent Neural Networks (RNNs), among others, have since been developed for images, and Natural Language Processing (NLP) applications, respectively. For the purpose of this study, we review RNNs and specifically the Seq2Seq architecture, widely used in NLP.    
\subsubsection{\color{black}{Temporal Neurons}}
RNNs were developed to preserve the temporal context of an input stream or sequence by introducing a weighted feedback line, called the state or memory, to the existing neuron structure. Each RNN cell takes in two inputs at a given time-step $t$ - (i) The input at that time-step $X_t$, and (ii) the cell state from the previous time-step $S_{t-1}$, to output the updated cell state $S_t$, and is given by
\begin{equation}
S_t = \Omega(WS_{t-1} + UX_t)
\label{RNN}
\end{equation}
where $W$, and $U$ are the learnable weights of the cell, and $\Omega$ is the non-linear activation function, typically tanh, Sigmoid or Rectified Linear Unit (ReLU). Similar to traditional NNs, the learning process is also carried out using backpropagation. However, due to the multiplicative nature of $S_{t-1}$ gradient terms (as $S_{t-1}$ comprises of nested $S_{t-2}, S_{t-3}, \dots, S_0$ terms using Eqn.~\ref{RNN}), RNNs suffer from a vanishing gradient problem as $t$ increases, resulting in failure to capture long term dependencies. To overcome this challenge, RNN variants such as Long Short Term Memory (LSTM) and Gated Recurrent Units (GRUs) came into existence \cite{goodfellow2016deep}. As GRUs are computationally efficient than LSTMs, with no considerable degradation in performance, they are the preferred choice of RNN in our study and are therefore discussed further. 
\par Similar to vanilla RNNs, GRUs also take in $X_t$ and $S_{t-1}$ as inputs to output $S_t$. However, the computation of $S_t$ occur in sub-stages, rather than Eqn.~\ref{RNN}. GRUs contain two additional computational units called (i) a read/reset gate, that outputs $R_t$, and (ii) a forget/update gate, that outputs $Z_t$. The read gate essentially reads some information from the previous state and the current input to compute an intermediate cell state $\tilde{S_t}$. The forget gate output $Z_t$ is then used to determine the information to update from $\tilde{S_t}$, and remove from $S_{t-1}$ to finally output the updated cell state $S_t$. The operations can be formally represented as
\begin{equation}
\begin{split}
R_{t} &= \sigma(W_rS_{t-1} + U_rX_{t}) \\
Z_{t} &= \sigma(W_zS_{t-1} + U_zX_{t}) \\
\tilde{S_t} &= \Omega(W(R_t \circ S_{t-1}) + UX_t) \\
S_{t} &= Z_t \circ S_{t-1} + (1-Z_t)\circ \tilde{S_t}
\end{split}
\label{GRU}
\end{equation}
\subsubsection{\color{black}{Basic Components of Seq2Seq Models}}\label{seq2seq-text}
Seq2Seq models are extensively used in the NLP domain for several applications ranging from machine translation, chatbots to keyword extraction \cite{lee2018scalable,yu2018multilingual,zhang2018keyphrase}. Our proposed approach is inspired specifically from the abstractive text summarization/ keyword extraction task, where the input would be a series of sentences (a paragraph) and the output would be a set of relevant keywords that do not necessarily be all present in the input sequence, as long as they are contextually relevant. Seq2Seq processing primarily occurs in three stages - viz. (i) Pre-processing, (ii) Encoding the input sequence, (iii) Decoding the desired output.

\begin{figure*}
\centering
\includegraphics[width=\textwidth]{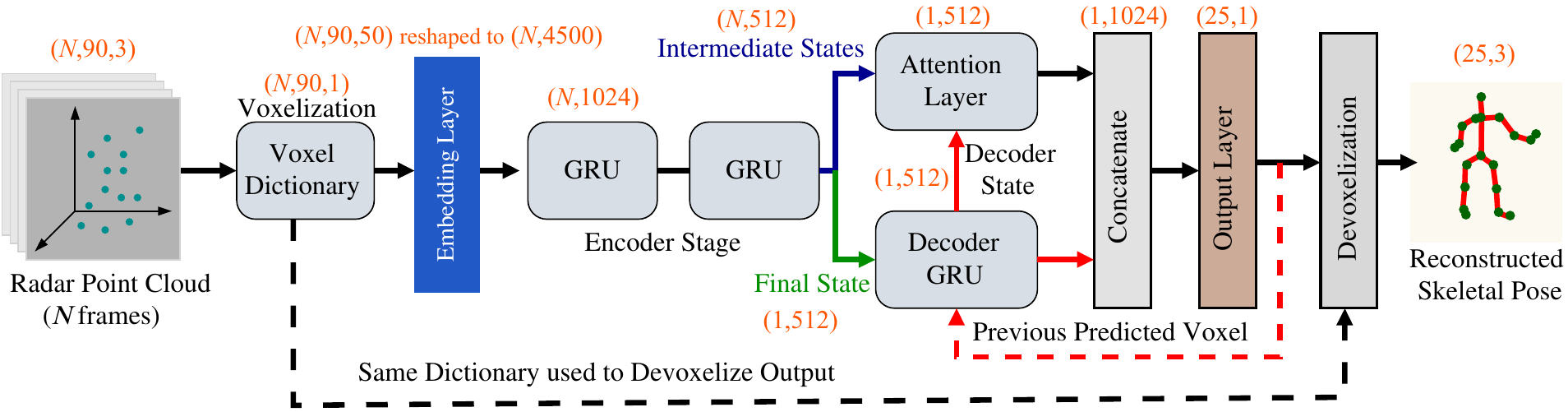}
\caption{The signal flow graph with dimensions (in red) of the proposed \textit{mmPose-NLP} architecture }
\vspace{-0.5cm}
\label{proposed}
\end{figure*}

\par Consider an input sentence of length $N$, represented as a sequence $\{x_1, x_2, \dots, x_N\}$, where every $x_i$ is a word in the vocabulary, defined as $V$. The first step in the pre-processing step is to assign a numeric value to each of the words, as NNs are essentially mathematical units and only work with numeric input-outputs. This is carried out using a process known as \textit{Tokenization}, where every words in the super-set vocabulary $V$ is mapped to an integer. This mapping is then used to convert the text database of input-outputs to a NN friendly numeric database. Besides tokenizing words from the vocabulary, two additional tokens - start-of-sequence $<sos>$ and end-of-sequence $<eos>$ - each respresented by unique integers are also added at the beginning and end of the output sequence, respectively. 
\par A naive approach to represent the tokenized input integer stream to the encoder stage of the model would be to use a one-hot representation. However, one-hot representation treats every word independently without capturing any similarity information among the words. A more prevalent approach, that also accounts for word similarity, is the use of an \textit{Embedding Layer}. The embedding layer is a trainable layer in the model that learns a mapping of the vocabulary to a higher ($M$) dimensional ``similarity" space, providing additional context to the NN. The similarity between two words, each represented as a $1 \times M$ embedded vectors $\tilde{x_1}$ and $\tilde{x_2}$, can be typically computed using a Cosine similarity metric, given by:
\begin{equation}
\cos \theta = \frac{\tilde{x_1} \cdot \tilde{x_2}}{||\tilde{x_1}||||\tilde{x_2}||}
\end{equation}
\par The tokenized data following the embedding layer is then sequentially fed into an \textit{Encoder}. The encoder comprises of several GRU units ($N_{enc}$), each operating independently and learning a unique encoded representation of the input sequence. The GRUs are first initialized with a cell state $EncS_{0,i}$, corresponding to the $i^{th}$ GRU. This cell state keeps updating with the relevant contextual information using Eqn.~\ref{GRU}, as the input comes in sequentially. The \textit{Decoder} also comprises of the same number of GRUs as the Encoder, and the initial cell states $DecS_{0,i}$ are initialized with the final cell-states of the GRUs from the encoder, i.e. $\{S_{N,1}, S_{N,2}, \dots, S_{N,N_{enc}}\}$. The $<sos>$ token serves as the first input to the decoder, and by using the initial decoder states (which are essentially the final encoded states), it sequentially predicts keywords, ending with the $<eos>$ token as output, once the prediction process is complete. Except for $<sos>$, the input to the decoder at a given time-step is the predicted keyword from the previous time-step.  Note that the the GRUs in the decoders only output their updated cell states at a given time-step, and not a direct keyword prediction. These cell states are typically subjected to a few fully connected layers leading upto the classification layer (with nodes the same size as the vocabulary), where a softmax activation determines the keyword that is predicted.  
\par In the vanilla Seq2Seq architecture, as described above, the decoder uses the final encoder state to initialize and start the prediction sequence. The attention layer, introduced in 2015, aims to overcome this approach of just relying on the final encoder state, but instead also looks at the intermediate encoder states, i.e. $\{S_{i,1}, S_{i,2}, \dots, S_{i,N_{enc}}\}$ $\forall i \in [0,N]$, to identify the amount of ``attention" to certain parts of the input sequence to aid the decoder make a prediction \cite{bahdanau2014neural, bahdanau2014neural2}. This is achieved in three stages. First an alignment score ($AS_{ij}$) is computed at every decoder prediction time-step $i$, to see how well the output at $i$ matches the context around the input time-step $j$ by using the decoder cell state $DecS_{i-1}$ and encoder cell state $EncS_{j}$. In the second stage, a softmax layer converts $AS_{ij}$ to a contextual probability distribution, yielding the attention weights $Att_{ij}$, given by:
\begin{equation}
Att_{ij} = \frac{e^{AS_{ij}}}{\sum_{k=1}^N e^{AS_{ik}}}
\end{equation}      
\par Finally, a context vector $CV_i$ is generated using a weighted sum of $Att_{ij}$ and $EncS_{j}$, given by $CV_i = \sum_{j=1}^N Att_{ij}EncS_{j}$. This context vector essentially aims to provide more emphasis or attention on certain portions of the input sequence and is then used in conjunction with the $DecS_{i-1}$ and the $(i-1)^{th}$ decoder prediction, to make the $i^{th}$ keyword prediction. 
\section{Proposed Approach}\label{proposed1}
The proposed \textit{mmPose-NLP} architecture has three main components - (i) Data acquisition, (ii) Data pre-processing, and (iii) Seq2Seq driven skeletal key-point estimator. For this study, we used two Texas Instruments (TI) AWR 1843 mmWave transceivers, each with 3 transmit (2 in azimuth and 1 elevation) and 4 receive antennas, at heights $h_1$, and $h_2$, and tilt angles $\theta_1$ and $\theta_2$, respectively, to sense a human subject in front of it. The rationale behind using two mmWave radars was to increase coverage in the elevation direction. The data from both the radars are acquired simultaneously and subjected to the radar signal processing chain (Section~\ref{RSPC-sec}). The channels in both azimuth and elevation allowed for a 3-D point cloud (PCL) representation of the target, yielding $[X_1^i,Y_1^i,Z_1^i]$ and $[X_2^j,Y_2^j,Z_2^j]$ in a given frame, where the subscript denotes the radar index (1/2), while the superscript ($i,j$) denote the 3-D coordinate of the $i^{th}$ and $j^{th}$ point in the PCL sets, respectively. Note that the 3-D coordinates, obtained using Eqn.~\ref{RSPC-eqn}, are with respect to the radars acting as the origin. However, as the radars are set at different heights, we translate the frame-of-reference to a common origin set at the ground plane. Also accounting for the tilt angles, the re-adjusted PCLs for the $k^{th}$ point from a radar index $n$ can be computed as:
\begin{equation}
\begin{split}
\tilde{X}_n^k &= X_n^k\cos\theta_n + Z_n^k\sin\theta_n\\
\tilde{Y}_n^k &= Y_n^k\\
\tilde{Z}_n^k &= Z_n^k\cos\theta_n - X_n^k\sin\theta_n + h_n\\
\end{split}
\label{corrected-eqn}
\end{equation} 
The re-adjusted PCLs $[\tilde{X}_1^i,\tilde{Y}_1^i,\tilde{Z}_1^i]$ and $[\tilde{X}_2^j,\tilde{Y}_2^j,\tilde{Z}_2^j]$ are then concatenated to form the total acquired and processed radar data set of PCL $[\tilde{X}_R,\tilde{Y}_R,\tilde{Z}_R]$, in a given frame. A Microsoft Kinect v2 also captures the scene, simultaneously with the radars, returning the 25 skeletal keypoints using the IR sensors $[X_k, Y_k, Z_k]$, that served as the ground truth (GT) or labels during the training process. 
\par Once the radar PCL and its corresponding GT labels are obtained, we used a Seq2Seq model to take in $N$ frames of the radar PCL as the input, and predict the desired skeletal key-points as output. Recollecting from Section~\ref{seq2seq-text}, such architectures are predominantly used in NLP applications that requires specific pre-processing. In order to achieve this, here are the analogous correlations of our proposed approach to the typical NLP application. The $N$ frames of the radar PCL is analogous to a paragraph of $N$ sentences. The desired 25 skeletal keypoints, correspond to the extracted keywords from the paragraph of text. A voxel dictionary is then created to map every 3-D point to a unique integer, analogous to \textit{Tokenization} in NLP. This is achieved by first identifying a cuboidal voxel space, centered at origin, with the voxel resolution of 5~cm (same as the radar resolution), in each of the 3 dimensions. The size of the cuboidal voxel space was determined empirically based on the maximum coverage of a human, centered at origin, performing various actions such as walking, stretching their arms sideways/above the shoulders, among others and was found to be 1.7m, 2.2m and 1.4m in the horizontal, vertical and depth axes, also defined as the region-of-interest (ROI). Also, to ensure that the that the voxelization is position invariant, i.e the point cloud from a subject is always inside the voxel space irrespective of the human’s distance from the radar, the 3-D centroid of the Radar and Kinect PCLs are first translated to the origin, and then the voxelization is performed. While the number of skeletal key-points obtained using a Kinect is fixed at 25, the number of points in the Radar PCL vary frame-to-frame. The maximum number of points, obtained empirically, in the radar PCL was found to be 88. To ensure constant input size, at each frame, the voxelized $M$ 3-D points in a radar PCL is always transformed to a fixed $1\times 90$ sized vector, where $90-M$ elements are set to a value of $0$, if $M<90$. The voxelized GT data results in a $1 \times 27$ vector, which represent the voxel indices of the 25 3-D skeletal keypoints, along with the start-of-sequence $<sos>$ and end-of-sequence $<eos>$ tokens added at the beginning and end of the sequence, respectively.
\par For the Seq2Seq model, $N$ frames of $1\times 90$ voxelized radar PCL data is first subjected to an $L$ dimensional embedding layer, resulting in an $N \times 90 \times L$ representation of the input. This data is then subjected to two layers of encoding, with $E_1$ and $E_2$ GRU units, respectively. Note that temporal neurons only accept two-dimensional data (time-step and feature vector). To allow for appropriate input dimension matching, the $N \times 90 \times L$ output from the embedding layer is reshaped to a $N \times 90*L$ tensor. The $N \times E_1$ output from the first encoder layer is fed into the second encoding layer that outputs (i) $N \times E_2$ tensor of all the intermediate cell states, and (ii) $1 \times E_2$ tensor of the final cell state. The decoder, comprising of $D = E_2$ GRUs, are initialized with the $1 \times E_2$ tensor of final encoder cell state as their initial state. Upon receipt of a $<sos>$ token, the decoder cell state, and the $N \times E_2$ intermediate cell states are fed into the \textit{Attention Layer}, where a $1 \times E_2$ context vector is generated. This context vector learns to provide emphasis on the parts of the input sequence that would aid in the prediction of the next skeletal key-point. This context vector along with the updated decoder cell state is concatenated and subjected to a fully connected layer for feature transformation followed by the output layer comprising of voxel indices, where the voxel index with the highest activation (using softmax) is predicted to be a keypoint. This keypoint is then fed back into the Decoder, and the process repeats until all the 25 skeletal key-points are predicted, upon which an $<eos>$ token output marks the end of the decoding process. The decoder \textit{always} predicts the key-points in a pre-defined order of joints, thus enabling the attention mechanism to learn the sections of the input to emphasize on, when predicting a specific key-point. Once the voxels of the 25 key-points are obtained, the voxel dictionary is used to convert them back to 3-D real world coordinates. The overall architecture is depicted in Fig.~\ref{proposed}.

\begin{table*}[h!]
\vspace{-0.25cm}
\centering
\caption{\textit{mmPose-NLP}'s Localization Accuracy Comparison}
\vspace{-0.25cm}
\label{MAE}
\resizebox{\textwidth}{!}{\begin{tabular}{c|
>{\columncolor[HTML]{EFEFEF}}c |
>{\columncolor[HTML]{EFEFEF}}c |
>{\columncolor[HTML]{EFEFEF}}c |
>{\columncolor[HTML]{C0C0C0}}c |
>{\columncolor[HTML]{C0C0C0}}c |
>{\columncolor[HTML]{C0C0C0}}c |
>{\columncolor[HTML]{EFEFEF}}c |
>{\columncolor[HTML]{EFEFEF}}c |
>{\columncolor[HTML]{EFEFEF}}c |
>{\columncolor[HTML]{C0C0C0}}c |
>{\columncolor[HTML]{C0C0C0}}c |
>{\columncolor[HTML]{C0C0C0}}c |}
\cline{2-13}
                                                                                                     & \multicolumn{6}{c|}{\cellcolor[HTML]{FFFFC7}Average MAE over 25 Skeletal Keypoints}                                                                                                                                                                                                        & \multicolumn{6}{c|}{\cellcolor[HTML]{FFFFC7}Average MAE over 17 Skeletal Keypoints}                                                                                                                                                                                                        \\ \cline{2-13} 
\multirow{-2}{*}{}                                                                                   & \multicolumn{3}{c|}{\cellcolor[HTML]{ECF4FF}\begin{tabular}[c]{@{}c@{}}Localization Error (cm)\\ {[}Random 20\% Test Split{]}\end{tabular}} & \multicolumn{3}{c|}{\cellcolor[HTML]{CBCEFB}\begin{tabular}[c]{@{}c@{}}Localization Error (cm)\\ {[}Continuous Test Dataset{]}\end{tabular}} & \multicolumn{3}{c|}{\cellcolor[HTML]{ECF4FF}\begin{tabular}[c]{@{}c@{}}Localization Error (cm)\\ {[}Random 20\% Test Split{]}\end{tabular}} & \multicolumn{3}{c|}{\cellcolor[HTML]{CBCEFB}\begin{tabular}[c]{@{}c@{}}Localization Error (cm)\\ {[}Continuous Test Dataset{]}\end{tabular}} \\ \hline
\multicolumn{1}{|c|}{\cellcolor[HTML]{000000}{\color[HTML]{FFFFFF} \textbf{Number of Input Frames}}} & \cellcolor[HTML]{ECF4FF}Depth              & \cellcolor[HTML]{ECF4FF}Horizontal              & \cellcolor[HTML]{ECF4FF}Vertical             & \cellcolor[HTML]{CBCEFB}Depth              & \cellcolor[HTML]{CBCEFB}Horizontal              & \cellcolor[HTML]{CBCEFB}Vertical              & \cellcolor[HTML]{ECF4FF}Depth              & \cellcolor[HTML]{ECF4FF}Horizontal              & \cellcolor[HTML]{ECF4FF}Vertical             & \cellcolor[HTML]{CBCEFB}Depth              & \cellcolor[HTML]{CBCEFB}Horizontal              & \cellcolor[HTML]{CBCEFB}Vertical              \\ \hline
\multicolumn{1}{|c|}{mmPose-NLP (1 Frames)}                                                          & 4.97                                       & 4.09                                            & 6.85                                         & 5.14                                       & 3.64                                            & 6.59                                          & 4.02                                       & 2.79                                            & 4.79                                         & 4.06                                       & 2.47                                            & 3.83                                          \\ \hline
\multicolumn{1}{|c|}{mmPose-NLP (2 Frames)}                                                          & \cellcolor[HTML]{EFEFEF}3.77               & \cellcolor[HTML]{EFEFEF}3.16                    & \cellcolor[HTML]{EFEFEF}4.16                 & 4.16                                       & 3.10                                            & 4.61                                          & 3.07                                       & 2.12                                            & 3.02                                         & 3.26                                       & 2.06                                            & 2.74                                          \\ \hline
\multicolumn{1}{|c|}{mmPose-NLP (3 Frames)}                                                          & \cellcolor[HTML]{EFEFEF}3.23               & \cellcolor[HTML]{EFEFEF}2.71                    & \cellcolor[HTML]{EFEFEF}3.29                 & 3.74                                       & 2.70                                            & 3.72                                          & 2.61                                       & 1.84                                            & 2.42                                         & 2.93                                       & 1.81                                            & 2.33                                          \\ \hline
\multicolumn{1}{|c|}{mmPose-NLP (4 Frames)}                                                          & \cellcolor[HTML]{EFEFEF}2.95               & \cellcolor[HTML]{EFEFEF}2.46                    & \cellcolor[HTML]{EFEFEF}2.92                 & 3.46                                       & 2.47                                            & 3.38                                          & 2.38                                       & 1.65                                            & 2.12                                         & 2.68                                       & 1.64                                            & 2.03                                          \\ \hline
\multicolumn{1}{|c|}{mmPose-NLP (5 Frames)}                                                          & 2.82                                       & 2.37                                            & 2.73                                         & 3.30                                       & 2.31                                            & 3.01                                          & 2.28                                       & 1.59                                            & 1.99                                         & 2.59                                       & 1.55                                            & 1.96                                          \\ \hline
\multicolumn{1}{|c|}{mmPose-NLP (6 Frames)}                                                          & 2.69                                       & 2.18                                            & 2.52                                         & 3.10                                       & 2.18                                            & 2.81                                          & 2.18                                       & 1.52                                            & 1.88                                         & 2.45                                       & 1.52                                            & 1.81                                          \\ \hline
\multicolumn{1}{|c|}{mmPose-NLP (7 Frames)}                                                          & 2.69                                       & 2.17                                            & 2.42                                         & 3.08                                       & 2.12                                            & 2.67                                          & 2.16                                       & 1.52                                            & 1.81                                         & 2.41                                       & 1.46                                            & 1.72                                          \\ \hline
\multicolumn{1}{|c|}{mmPose-NLP (8 Frames)}                                                          & 2.68                                       & 2.12                                            & 2.37                                         & 3.13                                       & 2.15                                            & 2.58                                          & 2.14                                       & 1.49                                            & 1.76                                         & 2.41                                       & 1.49                                            & 1.68                                          \\ \hline
\multicolumn{1}{|c|}{mmPose-NLP (9 Frames)}                                                          & 2.74                                       & 2.17                                            & 2.37                                         & 3.21                                       & 2.11                                            & 2.67                                          & 2.14                                       & 1.51                                            & 1.70                                         & 2.43                                       & 1.48                                            & 1.68                                          \\ \hline
\multicolumn{1}{|c|}{mmPose-NLP (10 Frames)}                                                         & 2.82                                       & 2.13                                            & 2.39                                         & 3.30                                       & 2.19                                            & 2.76                                          & 2.17                                       & 1.50                                            & 1.71                                         & 2.45                                       & 1.52                                            & 1.73                                          \\ \hline
\multicolumn{1}{|c|}{\textit{RF-Pose3D (8 Key-Points)}}                                                          & 4.2                                       & 4.9  & 4.0                                         & 4.2                                       & 4.9  & 4.0                                        
& 4.2                                       & 4.9  & 4.0                                        
& 4.2                                       & 4.9  & 4.0                                                                                  \\ \hline
\multicolumn{1}{|c|}{\textit{mm-Pose} (17 Key-Points)}                                                          & 3.2                                       & 7.5  & 2.7
& 3.2                                       & 7.5  & 2.7
& 3.2                                       & 7.5  & 2.7
& 3.2                                       & 7.5  & 2.7
 \\ \hline
\end{tabular}}
\end{table*}

\begin{figure*}
\includegraphics[width=\textwidth]{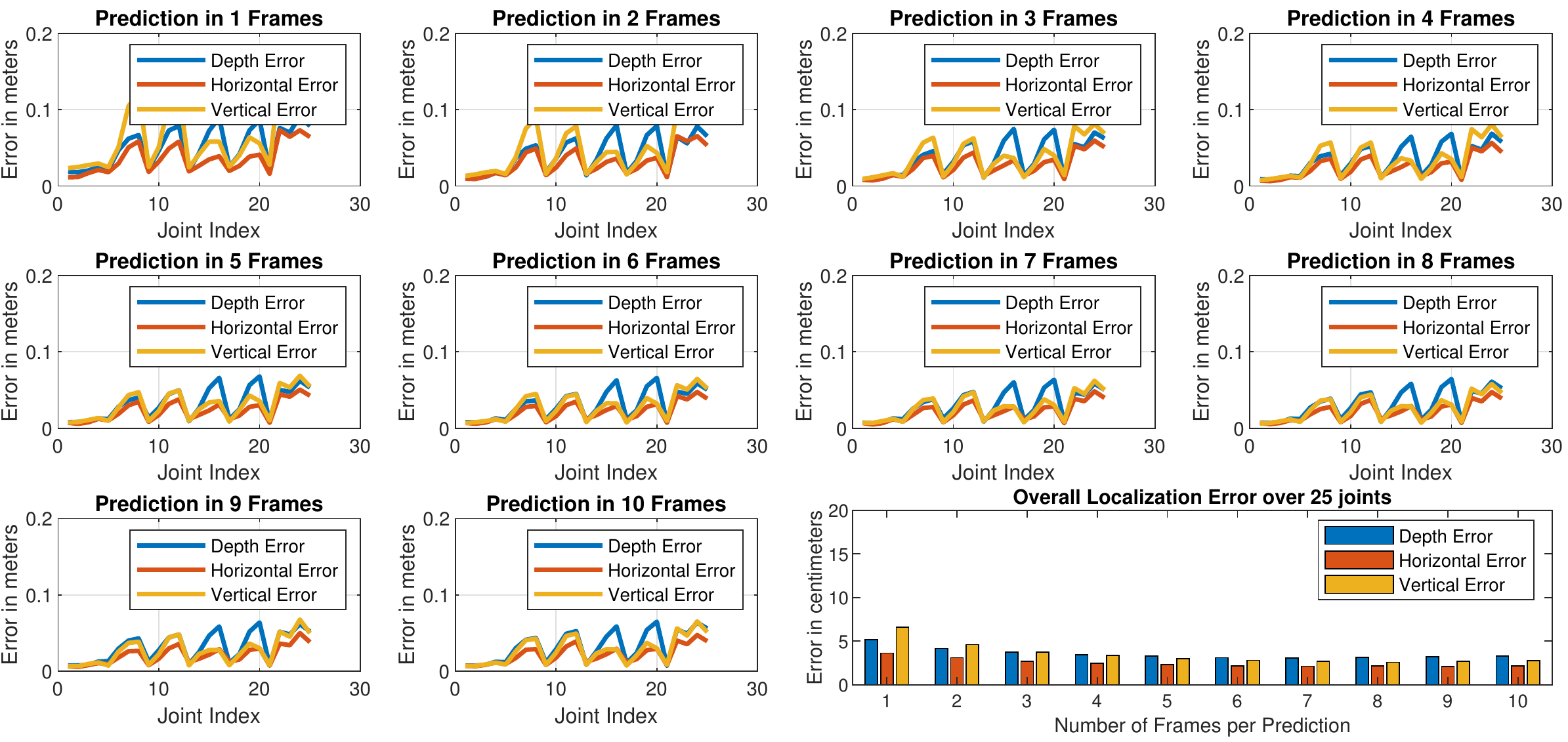}
\caption{Joint-wise averaged MAE for 25 skeletal key-points. }
\label{jntwise25}
\vspace{-0.5cm}
\end{figure*}
  
\section{Experimental Results and Discussion}\label{results}
\subsection{\color{black}{Setup and Data Collection}}\label{setup}
Two mmWave radars (top and bottom) and Kinect system was assembled and mounted on tripods at heights 2~m (top radar), 1~m (bottom radar), and 1.13~m (Kinect), and were setup in the living room of an apartment. The top and bottom radars were tilted at 20$^o$ and 15$^o$ in the depth-elevation plane, respectively. The radars were configured at start frequencies of 77 and 78 GHz, respectively, to avoid interference, with a chirp bandwidth of 3 GHz. Each radar frame or CPI constituted of 32 chirps, with the PRI (per channel) set to 90 $\mu$s, which equals to a total PRI of 270 $\mu$s for the three transmitters (using TDM). The processed 3-D radar PCL data (Eqn.~\ref{RSPC-eqn}) was acquired using USB interface on a robot operating system (ROS) interface, running on a Linux computer. Besides the 3-D PCL data, the radar packets also carried a header with the UTC time-stamp and the radar index (top or bottom). 
\par The GT data is aquired from the Kinect v2 sensor using a IR based skeleton keypoint tracker on a MATLAB API on a Windows computer. Similar to the radar PCL frames, the 25 skeletal key-points from the Kinect is also accompanied by a UTC time-stamp, in every frame. The Linux and Windows computers are synchronized using a common time-server. These UTC stamps are then used for frame association and data-set formulation. We collected a total of 15000 frames of data with two human subjects performing a series of actions ranging from walking, left arm swing, right arm swing, and both arms swing. The height-offsets and tilt-angle corrections were applied (Eqn.~\ref{corrected-eqn}) to obtain the 3-D PCLs in a unified frame of reference. The corrected 3-D radar PCL and 3-D GT key-point PCL were then used to generate the training and test datasets, as discussed in the next section.

\begin{figure*}
\includegraphics[width=\textwidth]{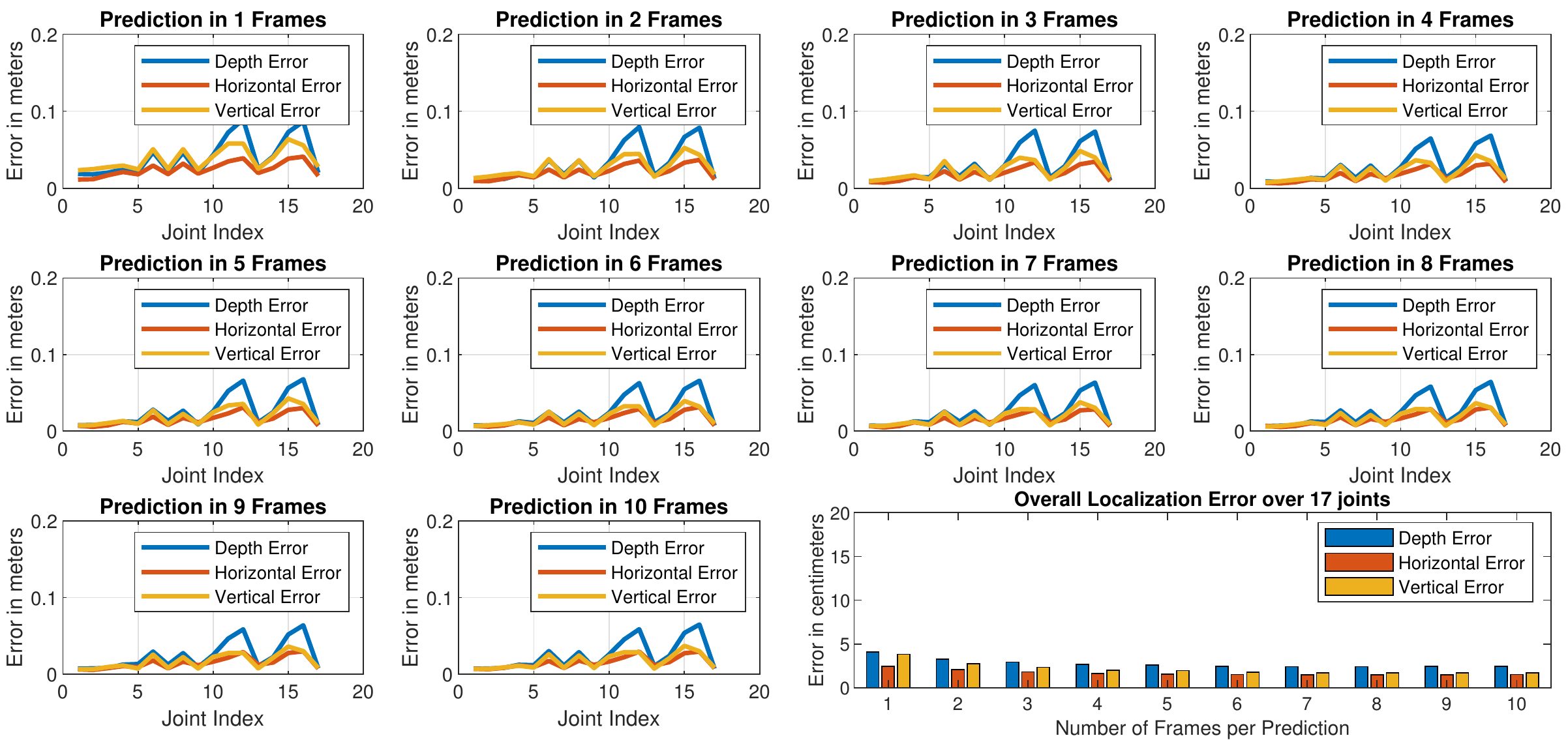}
\caption{Joint-wise averaged MAE for 17 skeletal key-points after outlier detection.}
\label{jntwise17}
\end{figure*}

\begin{figure}
\includegraphics[width=0.5\textwidth]{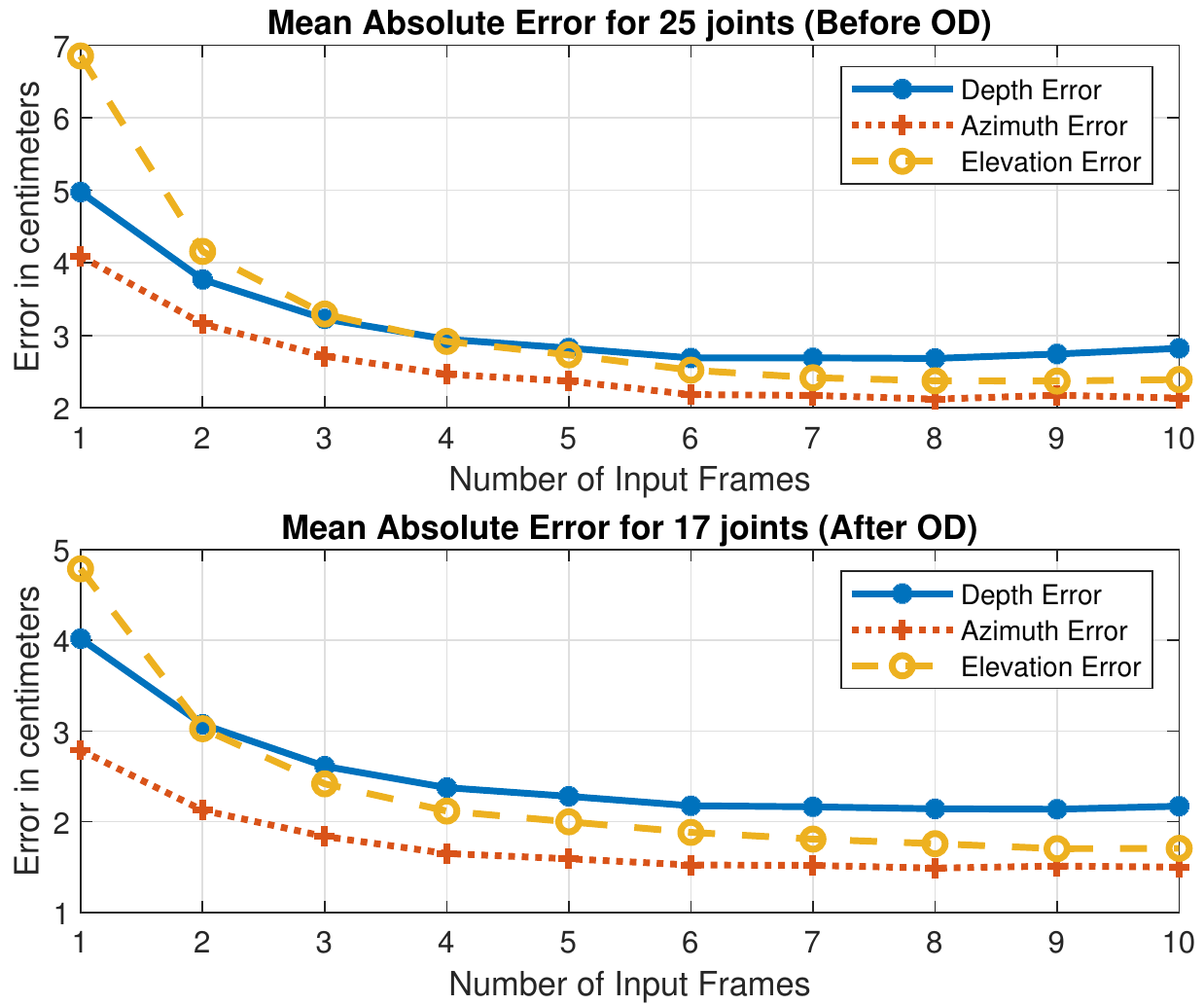}
\caption{The localization MAE trend with respect to the number of input radar frames  for 25 and 17 skeletal keypoints, respectively.}
\label{MAE25-17}
\end{figure}

\begin{figure*}
\vspace{-0.5cm}
\centering
\includegraphics[width=0.9\textwidth]{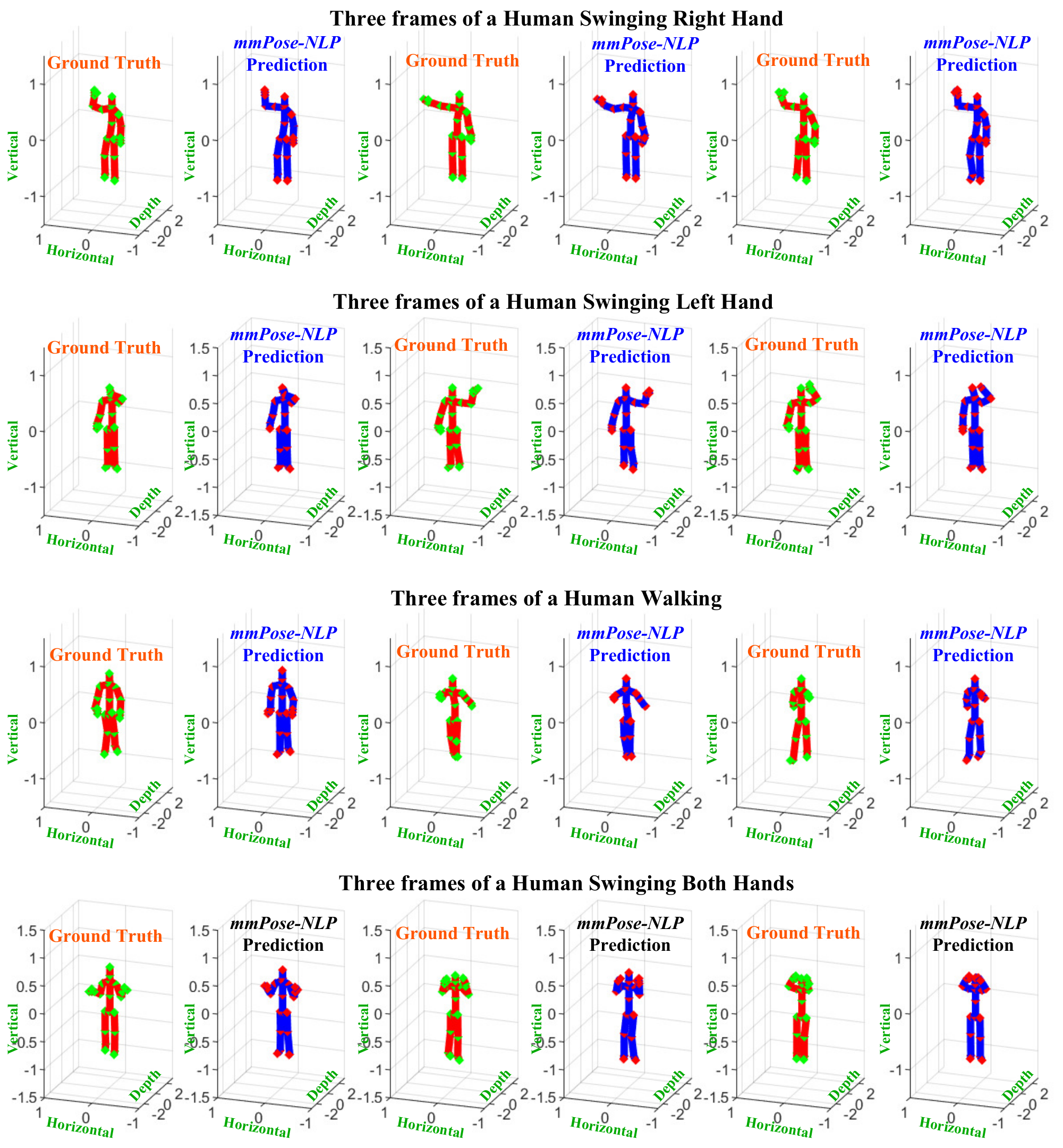}
\caption{Pictorial comparison of three frames of the Ground Truth skeletal pose vs \textit{mmPose-NLP} predicted skeletal pose for 4 different actions.}
\label{skeleton}
\end{figure*}

\begin{figure}
\includegraphics[width=0.5\textwidth]{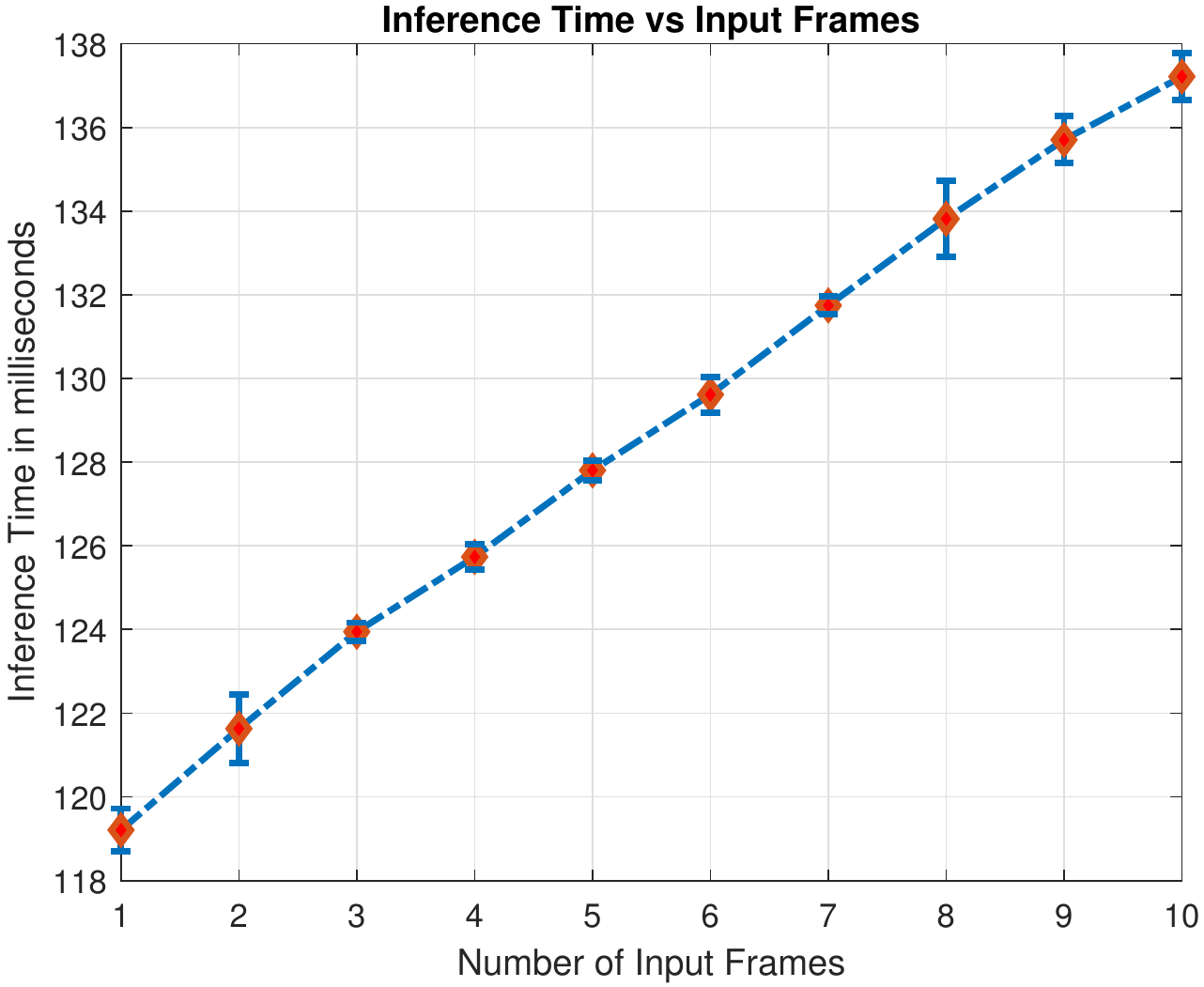}
\caption{The mean inference time to extract 25 skeletal key-points compared against the number of input frames.}
\vspace{-0.5cm}
\label{time}
\end{figure}
\subsection{\color{black}{Training mmPose-NLP}}
Prior to training the \textit{mmPose-NLP} model, the 3-D radar and GT key-point data were tokenized using the voxelization approach, outlined in Section.~\ref{proposed1}. Then $N$ frames of the voxelized radar data were sequentially combined to result in a $N \times 90$ input tensor. In this study, we evaluate the proposed \textit{mmPose-NLP} model for $N = \{1,2,\dots,10\}$, therefore 10 distinct data-sets ($Data_{tot_N}$) were generated. For the output labels (voxelized GT data), $<sos>$ and $<eos>$  tokens were added at the beginning and at the end of the sequence for each of the 10 data-sets. A 1200 frame long continuous sequence was first branched out from the data-set as our first test data-set ($Test_{cnt}$), mainly to aid visual interpretation of the model performance and comparison.
\par The proposed \textit{mmPose-NLP} Seq2Seq model as presented in Section.~\ref{proposed1} was implemented using Keras Functional API on a Tensorflow v2 backend. Except for the input dimensions ($N \times 90$) that depended on the number of frames, that in turn governed the output size from the 50 dimensional embedding layer ($N \times 4500$), the remainder of neurons in the network remained the same for each of the ten $N$-frame input scenarios. For each $N$-frame case, the remainder of the data-set ($Data_{tot_N} - Test_{cnt}$) were shuffled, and 5 distinct $Train_{ij}$ and $Test_{ij}$ data-sets (80\% - 20\% split) were generated, similar to a five-fold cross validation approach, and the the test data-sets were saved to disk for model evaluation. The models $M_{ij}$, where $i$ is the frame index $\in \{1,2,\dots,10\}$, for the $j^{th}$ training data-set $Train_{ij}$ $j \in \{1,2,\dots,5\}$. The models were trained on UArizona's High Performance Computer (HPC) clusters that employ NVidia P100 GPUs, with the objective to minimize the sparse categorical cross entropy. The use of sparse categorical cross entropy alleviates the need to further convert the output labels to a one-hot-vector representation which is memory intensive. Upon training, the models were saved to disk for offline evaluation on the test data-sets.
\subsection{\color{black}{Results and Discussion}}
The trained models $M_{ij}$ were then subjected to the test datasets - (i) The continuous test data-set $Test_{cnt}$ that was branched out initially; and (ii) The shuffled and 20\% test data split $Test_{ij}$, that were saved to disk. The predicted voxels from the test data were converted back to real-world 3-D coordinates using the same voxel dictionary that was used during the tokenization process.

\subsubsection{\color{black}{Localization Accuracy}}
The predicted 25 skeletal key-points ($Pred_{ij}$) are compared against the 25 ground truth labels ($GT_{ij}$) for accuracy in depth, vertical and horizontal directions by computing the Mean Absolute Error ($MAE_{ij}$). Furthermore, the MAE for every $N$-frame scenario is obtained using the average MAE across the five test instances, i.e, $MAE_i = \frac{1}{5}\sum_{j=1}^5 MAE_{ij}$, $\forall i\in \{1,2,\dots,10\} $. MAE was also computed for ($Test_{cnt}$) over all the $10 \times 5$ models and were averaged to obtain the mean MAE for each of the $1$ to $10$ frame scenarios. Consistent with our previous findings, the localization errors were the highest for 8 joints, namely the wrist, index finger, palm center, and the thumb of both left and right hands. While the
ground truth data using Kinect could resolve for these joints on account of dense IR point cloud, representing small RCS joints using mmWave radar returns alone could be quite challenging. However, despite a comparatively higher error on these 8 joints, the overall localization MAE for the 25 joints provide significant improvement over the current approaches, viz. \textit{RF-Pose} (8 key-points) and \textit{mm-Pose} (17 key-points). We also performed an outlier detection (OD) based MAE by removing the 8 outlier key-points for a further comparison with our previous work \textit{mm-Pose}, that computed localization MAE with the remainder of 17 joints. The localization MAE results for all the $N$ frame cases have been outlined in Table.~\ref{MAE}. The joint-wise MAE for the 25 joints (before OD) and 17 joints (after OD) have been presented in Fig.~\ref{jntwise25} and Fig.~\ref{jntwise17}, respectively. 
\par In this study, we also looked at the effect of the number of input frames on the overall accuracy, or in this case the MAE. The primary objective of the proposed approach is to exploit and use temporal information from the consecutive frames of the radar data to determine the position of the skeletal key-points. In our study, we observed a gradual improvement in the MAE with an increase in the number of input frames from 1 through 8 frames, consistent with our hypothesis. However, we saw a slight increase in the MAE as the number of frames as input rose to 8 through 10. This is depicted in Fig.~\ref{MAE25-17}. This implies that for the current radar configuration parameters, setup, and the proposed \textit{mmPose-NLP} architecture, $N=8$ offers the best localization accuracy. Finally, a pictorial comparison of the \textit{mmPose-NLP} generated skeletal pose, compared against the GT for walking, left swing, right swing and both swing has been presented in Fig.~\ref{skeleton}. Animated comparison of \textit{mmPose-NLP} predictions vs GT on $Test_{cnt_N}$ for all the $N \in \{1,2,\dots,10\}$ frame cases, along with the source codes for this project is available on \mbox{\textit{https://github.com/radar-lab/mmPose-NLP}}.  
\subsubsection{\color{black}{Limitations}}
While the proposed Seq2Seq inspired \textit{mmPose-NLP} architecture offers a better localization accuracy over previously proposed radar-based pose estimation techniques, the iterative nature of the key-point prediction is time-intensive, as shown in Fig.~\ref{time}, making it unsuitable for real-time skeletal key-point estimation at the moment. Although the proposed approach could find several key applications that require highly-accurate offline skeletal pose estimation such as Biomechanics (Gait Kinematics) and Defense Security Forensics, the authors aim to continue investigating techniques to alleviate the inference time to also aid several critical real-time applications such as autonomous vehicles, traffic monitoring and automated patient monitoring. Also, as with any supervised technique, the current architecture has only been verified with the data format, resolution and setup (height/tilt angles) as specified in Section~\ref{setup}. While the underlying methodology would still hold for similar application, the neural network parameters would need adjusting/fine-tuning for optimum feature extraction from data obtained using a different set of configuration parameters.    
\section{Conclusion}\label{conclusion}
In this paper, we proposed \textit{mmPose-NLP}, an NLP inspired Seq2Seq based skeletal pose estimator using mmWave radar data. The radar point cloud data was first voxelized and $N$ consecutive frames were fed into the \textit{mmPose-NLP} Seq2Seq model. The predicted key-point voxels were converted back to the real world 3-D coordinates using the voxel dictionary. Data was collected with two AWR 1843 mmWave radars on ROS, and the ground truth was obtained using Kinect's skeletal tracker on MATLAB. Two human subjects performed a four continuous actions, namely (i) Walking, (ii) Left Hand Swing, (iii) Right Hand Swing, and (iv) Both Hands Swing. Besides a separate continuous test data-set, each $N$ frame input scenario consisted of five distinct shuffled 80\%-20\% training-test splits from the data-set for comprehensive evaluation. The temporal information and the key-point specific emphasis via the attention layers allowed the proposed approach to achieve accurate reconstruction of a skeletal pose with $<$3~cm localization errors in the depth, azimuth and elevation axes. To the best of the authors knowledge, the proposed approach is the first method to reliably estimate upto 25 skeletal key-points, with an improved accuracy over the existing contemporaries. Furthermore, a study of the localization accuracy vs the number of input frames was carried out and presented. The proposed approach can find several key applications in forensic and clinical gait analysis, patient monitoring systems, human computer interaction, and the emerging AI-powered fitness systems, especially with mmWave radars not capturing facial details, given the increasing concerns with user privacy. With continued efforts to reduce the inference time, the proposed approach could find applications in autonomous vehicles and traffic monitoring systems for timely decision making. 
\section*{Acknowledgements}
The authors would like to thank Aditi Deshpande for volunteering and helping with the data collection process, amidst the challenging COVID-19 restrictions.
\bibliographystyle{ieeetr}
\bibliography{references}

\end{document}